\documentclass[%
 reprint,
superscriptaddress,
 amsmath,amssymb,
aps,
prb,
10pt
]{revtex4-1}

\usepackage[utf8]{inputenc}
\usepackage{microtype}
\usepackage{graphicx}
\usepackage{dcolumn}
\usepackage{bm}
\usepackage{verbatim}
\usepackage{booktabs}
\usepackage{mathptmx} 

\begin{document}

\title{Modulator-free coherent-one-way quantum key distribution}

\author{George L. Roberts}
\affiliation{Toshiba Research Europe Ltd, 208 Cambridge Science Park, Milton Road, \\Cambridge CB4 0GZ, United Kingdom}%
\affiliation{Cambridge University Engineering Department, 9 J J Thomson Avenue, \\Cambridge, CB3 0FA, United Kingdom}%
\author{Marco Lucamarini}%
\affiliation{Toshiba Research Europe Ltd, 208 Cambridge Science Park, Milton Road, \\Cambridge CB4 0GZ, United Kingdom}%
\author{James F. Dynes}%
\affiliation{Toshiba Research Europe Ltd, 208 Cambridge Science Park, Milton Road, \\Cambridge CB4 0GZ, United Kingdom}%
\author{Seb J. Savory}%
\affiliation{Cambridge University Engineering Department, 9 J J Thomson Avenue, \\Cambridge, CB3 0FA, United Kingdom}
\author{\\Zhiliang Yuan}%
\email{zhiliang.yuan@crl.toshiba.co.uk}
\affiliation{Toshiba Research Europe Ltd, 208 Cambridge Science Park, Milton Road, \\Cambridge CB4 0GZ, United Kingdom}%
 \author{Andrew J. Shields}%
\affiliation{Toshiba Research Europe Ltd, 208 Cambridge Science Park, Milton Road, \\Cambridge CB4 0GZ, United Kingdom}%

\begin{abstract}
Time-bin encoding is an attractive method for transmitting photonic qubits over long distances with minimal decoherence. 
It allows a simple receiver for quantum key distribution (QKD) that extracts a key by measuring time of arrival of photons and detects eavesdropping by measuring interference of pulses in different time bins. 
In the past, coherent pulses have been generated using a CW laser and an intensity modulator. 
A greatly simplified transmitter is proposed and demonstrated here that works by directly modulating the laser diode. 
Coherence between pulses is maintained by a weak seed laser. 
The modulator-free source creates time-bin encoded pulses with a high extinction ratio (29.4~dB) and an interference visibility above 97~\%. 
The resulting QKD transmitter gives estimated secure key rates up to 4.57~Mbit/s, the highest yet reported for coherent-one-way QKD, and can be programmed for all protocols using weak coherent pulses.
\end{abstract}

\maketitle

Quantum key distribution (QKD) uniquely allows two parties to exchange secure keys with secrecy guaranteed by the fundamental laws of physics~\cite{gisin_quantum_2002}. 
Its potential for real-world applications has stimulated a large amount of progress in developing implementation technologies~\cite{Diamanti_practical_2016}. 
Over optical fiber links, QKD has been demonstrated to distribute quantum keys with rates exceeding 1~Mbit/s~\cite{dixon_continuous_2010}, over a hundred kilometers of distance~\cite{Korzh_provably_2015,frohlich_long-distance_2017} and/or in the presence of strong classical signals in the same fiber~\cite{patel_quantum_2014}. 
The technology is being extensively tested in installed fiber network environments~\cite{Qiu_quantum_2014}.
Moreover, satellite QKD and quantum repeater technologies are also being pursued to extend communication to the global scale~\cite{Diamanti_practical_2016}. 

Since the inception of QKD in 1984 with the Bennett-Brassard (BB84) protocol~\cite{Bennett_Quantum_1984}, a number of diverse implementations have been proposed. 
Two broad classes of protocol that share popularity are discrete variable and distributed phase reference protocols~\cite{Scarani_security_2009}. 
The protocols within these classes have a variety of requirements for transmitters, for example phase modulation and/or intensity modulation. 
Each provides different benefits, whether it is high bit rate, long achievable distance, a rigorous security proof or simplicity in experimental implementation. 
It is therefore highly beneficial to develop a versatile QKD transmitter that can operate different QKD protocols.  
Promising work has recently been demonstrated by combining a laser with a number of external phase and intensity modulating elements~\cite{sibson_integrated_2017}. 

The modulator-free transmitter proposed by Yuan \textit{et al}~\cite{yuan_directly_2016} has a number of attractive properties for phase modulation. 
The light source uses a pair of laser diodes in an optical injection configuration. 
A master laser provides phase modulation, and randomization when required, while the slave laser is responsible for generating short optical pulses.
The light source has successfully been demonstrated for two important phase-encoded QKD protocols, BB84 and differential phase shift (DPS)~\cite{Inoue_differential_2002}. 
However, the suitability of a modulator-free design to offer high extinction ratio intensity modulation in quantum communications is yet to be explored. 

Here, we tackle this issue by implementing the coherent-one-way (COW) QKD protocol with a modulator-free transmitter. 
This transmitter allows us to achieve record-breaking key rates with quantum bit error rates (QBERs) below 1~\% and visibilities over 97~\%. 
Using a realistic finite key-size scenario, we can distribute keys from Alice to Bob at losses between 1.5~dB and 30~dB, equivalent to 7.5~km and 150~km of standard single mode optical fiber.

The COW protocol~\cite{Korzh_provably_2015} uses time-bin encoding to share a key between two parties. 
Security of the key is ensured by Alice maintaining a fixed coherence between pulses. 
Bob can measure the interference visibility between adjacent pulses using an interferometer and infer the presence of an eavesdropper, Eve, by a break in the coherence. 
It has been used in a real-fiber system to transmit a secure key between two parties separated by 307 km - the longest distance for any two party quantum protocol~\cite{Korzh_provably_2015}. 
This is possible because the time-bin encoding produces a lower QBER than other protocols, for example BB84~\cite{dixon_continuous_2010}. 
The main downside to the protocol is that whilst security against a number of attacks has been demonstrated, no comprehensive security proof exists for all families of attack~\cite{Scarani_security_2009}, which could mean the protocol is vulnerable to an all-powerful Eve. 

In the COW protocol, Alice prepares two values of a logical bit using empty or full time bins: $\rvert$$\beta_0$$\rangle$=$\rvert$$\alpha$$\rangle$$\rvert$0$\rangle$ and $\rvert$$\beta_1$$\rangle$=$\rvert$0$\rangle$$\rvert$$\alpha$$\rangle$, where $\rvert$0$\rangle$ is the vacuum state and $\rvert\alpha\rangle$ represents a coherent state of light~\cite{loudon_quantum_2000} with intensity $\mu$=$|\alpha|^2$. 
For these requirements, the transmitter used in this protocol must be able to modulate intensity whilst maintaining coherence. 
Bob decodes the signals by measuring their arrival times with a single photon detector (SPD). 
He also takes a portion of the received photons in order to measure the coherence between adjacent time bins, allowing him to test the channel against unauthorized external intrusions. 
This measurement is performed by overlapping two consecutive optical pulses on a beam splitter and measuring the resultant interference visibility. 
This is possible for the pulse sequence $\rvert$$\beta_1$$\rangle$$\rvert$$\beta_0$$\rangle$. 
To increase the number of consecutive non-empty pulses, Alice also prepares a decoy sequence $\rvert$$\beta_2$$\rangle$=$\rvert$$\alpha$$\rangle$$\rvert$$\alpha$$\rangle$. 
This reduces the amount of bits Bob needs to collect before he can accurately measure the visibility, which is important in the finite key-size scenario. 
Moreover, the decoy sequence is used as a security feature, as Eve does not know whether she is attacking a logical bit or a decoy sequence. 
During sifting, Alice informs Bob when she sent decoy pulses and Bob tells Alice whether he measured the arrival time or the visibility of the optical pulses. 

\begin{figure}[ht!]
\centering
\includegraphics[width=\linewidth]{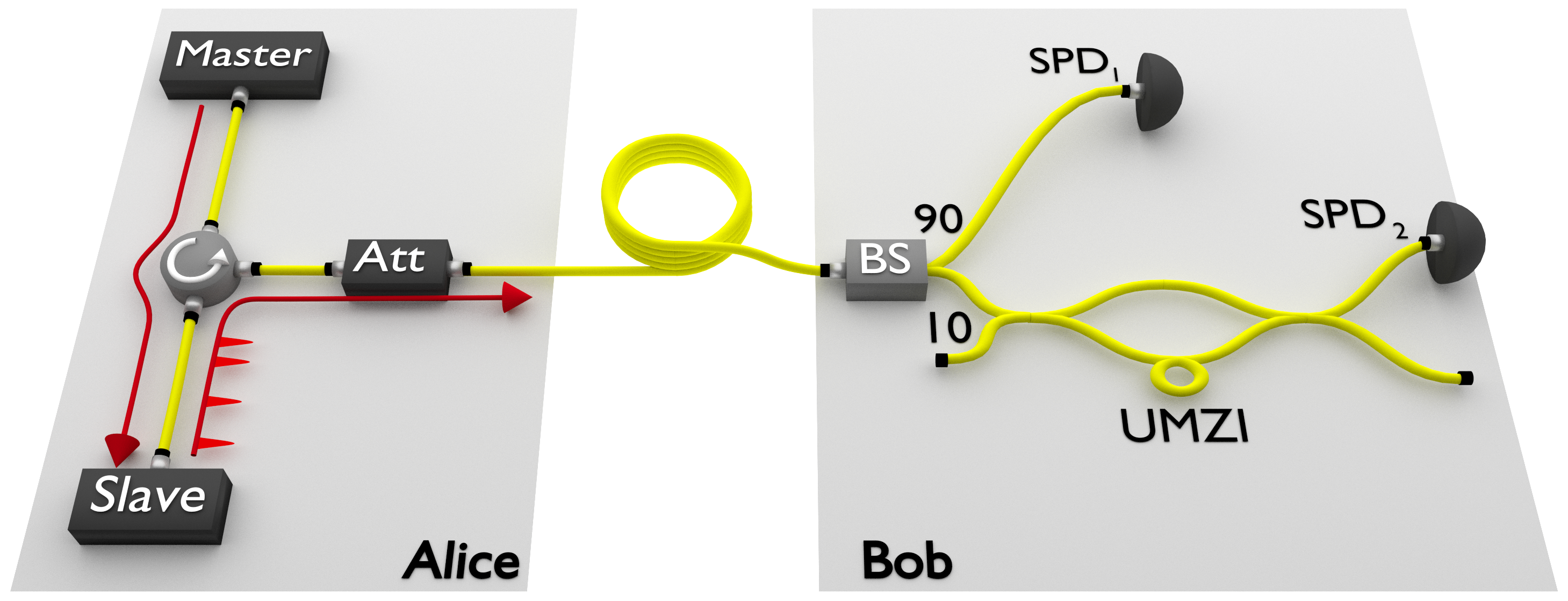}
\caption{\textbf{System schematics}. The master laser in Alice injects CW light into the slave laser, which produces low-jitter gain-switched pulses. These are then attenuated to the single photon level before being transmitted through the quantum channel to Bob. SPD$_1$ detects the arrival time of the photons, from which the key is generated, and SPD$_2$ is used to measure the phase coherence. BS=beamsplitter; UMZI=unbalanced Mach-Zehnder interferometer; att=attenuator.}
\label{fig:transmitter}
\end{figure}

Figure~\ref{fig:transmitter} shows the experimental setup.  
The quantum transmitter (Alice) is configured for time-bin encoding using a slave laser that is optically seeded by a master laser. 
An optical attenuator attenuates the non-empty time bins to around 0.1~photons per pulse before transmitting them through the quantum channel, implemented by an optical attenuator, to the quantum receiver (Bob).  
Bob uses a 90:10 beamsplitter to passively route most of the photons to a superconducting nanowire detector (SPD$_1$) for arrival time measurements. 
The remaining 10~\% are fed into an unbalanced Mach-Zehnder interferometer (UMZI) for measuring the phase coherence with a second superconducting nanowire detector (SPD$_2$). 
The UMZI is based on a planar lightwave circuit with a differential delay of 500~ps and has a loss of 3~dB.  
A built-in heater allows direct control of the phase delay across one arm.
The superconducting nanowire SPDs used feature a dark count rate (DCR) of 10 Hz, alongside an efficiency of 34~\% at a wavelength of 1550~nm, allowing us to reach long distances.

Intensity-modulated gain-switched pulses are produced through electrical modulation of the slave laser at 3.3~V~\cite{liu_Modulator-free_2014,He_amplification_2016}. 
A DC bias above the lasing threshold is applied to the master laser to ensure the phase is coherent when it is injected into the slave laser. 
We use a wavelength-tunable, continuous-wave fiber laser as the master laser and a semiconductor distributed feedback (DFB) laser diode as the slave. 
The slave laser is kept at room temperature with a free-running wavelength of 1550.1~nm.  
The master laser is wavelength-tuned to give a maximum coherence transfer, which occurs when both lasers have the same free-running wavelength. 
The output pulses from this system have a pulse width of 70~ps, which is much smaller than the inverse modulation frequency, and a spectral width of 0.10~nm. 
For time-bin encoding, a pseudo-random number generator creates a repeated 512-bit sequence, generating decoy sequences with a probability of 1~\% and signal sequences for the remaining time bins equally distributed between the bit values 0 and 1. 
The pattern generated is applied to gain-switch the slave laser at a clock rate of 2~GHz, therefore implementing a COW transmitter at an effective bit-rate of 1~GHz. 
A high intensity extinction ratio of 29.4~dB can be achieved between non-empty and empty pulses, thus ensuring a low encoding error in the time basis.

Optical injection ensures there is coherence among the gain-switched slave pulses, as they all inherit the phase of the CW master laser. 
The injected light transfers the coherence, thereby suppressing the randomness of the phase that would occur if pulses were triggered by spontaneous emission~\cite{Comandar_quantum_2016}. 
To illustrate the physical principle, we gain-switch the slave laser to produce a 2~GHz pulse train and replace the SPD$_2$ in Fig.~\ref{fig:transmitter} with an optical power meter to measure the interference fringe visibility, defined as
\begin{equation}
\label{eq:IC}
V=\frac{I_{max}-I_{min}}{I_{max}+I_{min}}
\end{equation}
\noindent where I$_{max}$ and I$_{min}$ are the average pulse intensities for constructive and destructive interference respectively.
The attenuator in Fig.~\ref{fig:transmitter} is set to maximum transmission, while a second attenuator (not shown) is used to vary the seed optical power into the slave laser. 
The interference visibility increases monotonically with seed power, as a result of the increasing dominance of the injected light over spontaneous emission in the slave laser cavity. 
The visibility saturates at 99.78~\% with a seed power of 216~$\mu$W.
In order to achieve a visibility of 99~\%, a modest 12~$\mu$W of seed power is sufficient.
In the subsequent QKD experiment, we use a seed power of 50~$\mu$W to ensure the pulses are sub-100~ps. 

\begin{figure}[h!]
\centering
\includegraphics[width=0.98\linewidth]{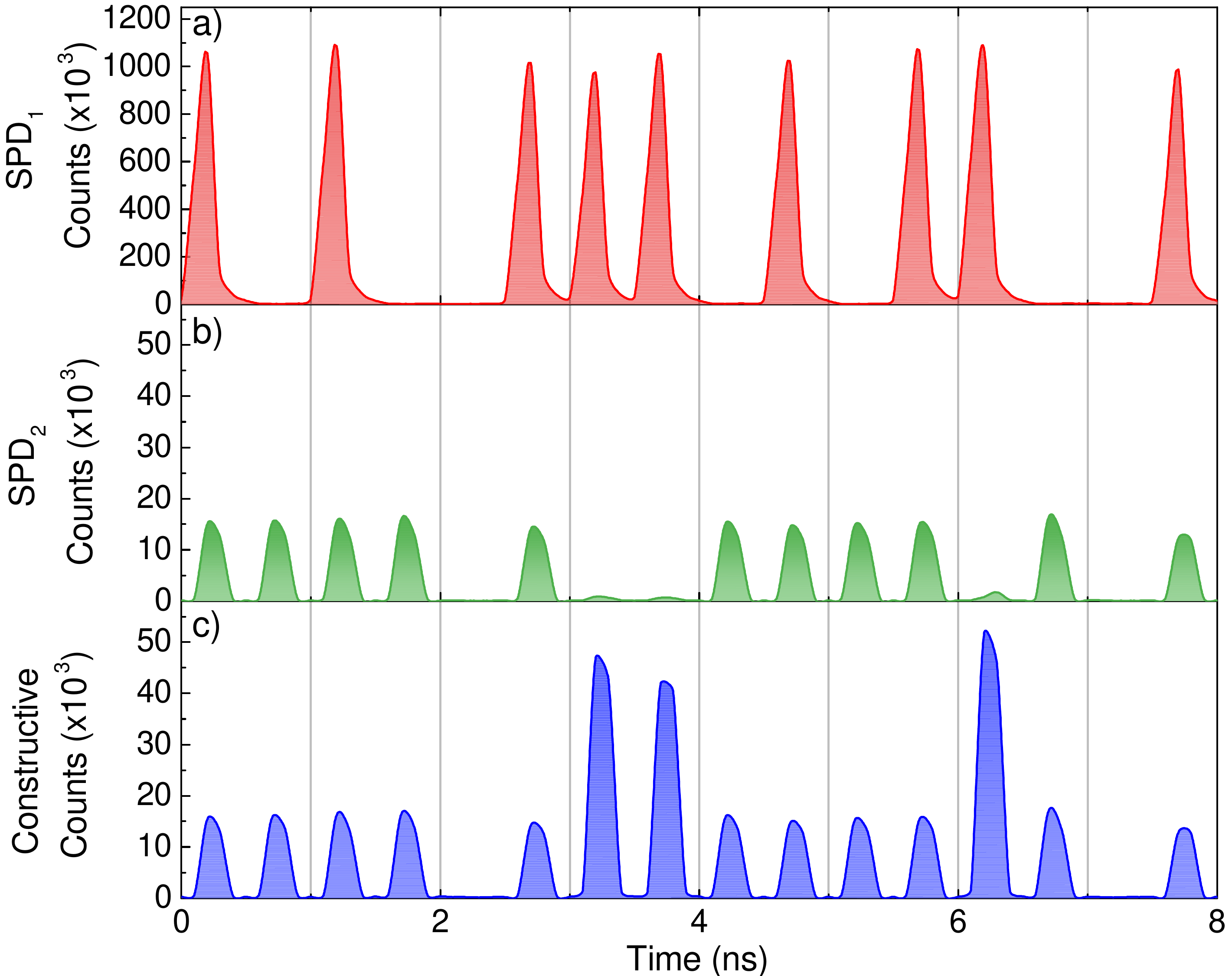}
\caption{\textbf{Detected signals.} Complementary signals received by Bob in a) the time basis; b) the destructive arm of his interferometer; c) the constructive arm of his interferometer, measured using the SPDs. The transmitted key for these patterns is $\rvert$$\beta_0$$\rangle$, $\rvert$$\beta_0$$\rangle$, $\rvert$$\beta_1$$\rangle$, $\rvert$$\beta_2$$\rangle$, $\rvert$$\beta_1$$\rangle$, $\rvert$$\beta_1$$\rangle$, $\rvert$$\beta_0$$\rangle$, $\rvert$$\beta_1$$\rangle$, where each logical bit is separated by a vertical grey line. Data is acquired for 60~s in a quantum channel with 15~dB loss, with Alice transmitting 0.1 photons per pulse.}
\label{fig:oscPatterns}
\end{figure}

The quantum transmitter and receiver are linked via a short optical fiber and extra attenuation is applied to simulate the loss of the quantum channel.
The transmitting photon flux is set to 0.1 photons per non-empty pulse at the output of the transmitter. 
At the lowest attenuation we decrease the photon flux to 0.07 photons per pulse to minimize time-jitter effects, as described later. 
In the QKD experiment, two channels of a digitizer with 100~ps time resolution simultaneously record the arrival times of single photons at the photon detectors (SPD$_1$ and SPD$_2$). 
Bob uses a 90:10 beamsplitter to passively direct most of the photons to SPD$_1$, where he sifts the key by measuring time-bins. 
The sifting loss here is minimal, caused only by the small portion of photons routed through the phase coherence measurement path. 
An example histogram measured by SPD$_1$ is shown in Fig.~\ref{fig:oscPatterns}(a), giving a QBER of less than 1~\%.
SPD$_2$ is placed at the destructive output port of the interferometer to enable an accurate measurement of the visibility.
To highlight the interference effect, we show in Fig.~\ref{fig:oscPatterns}(c) an example measurement from the constructive output port of the interferometer.
The height of constructive peaks is approximately four times that of the non-interfering peaks, as expected from first-order optical interference. 

Each QKD session is continued until over 2$\times$10$^7$ counts are collected in the time basis.
The QBER and visibility are collected alongside the number of counts in each arm.
The key rates are calculated using the finite key size analysis derived by Korzh \textit{et al}~\cite{Korzh_provably_2015} with a total security parameter of $\varepsilon_{QKD}=10^{-10}$. 
The key rate dependence on visibility and number of photons per pulse, $\mu$, is given by
\begin{equation}
\begin{split}
\zeta &= (2V-1)\times \exp(-\mu) \\ & 
-2\left\{\left[1-\exp(-2\mu)\right]V(1-V)\right\}^{1/2}. 
\end{split}
\end{equation}
The extracted key length is then calculated using 
\begin{equation}
\begin{split}
l &= n \left[1-Q-(1-Q)\textrm{h}\left(\frac{1-\zeta}{2}\right)\right] - 7\left[n \log_2(\beta^{-1})\right]^{1/2} \\&
- f_{IR}\times \textrm{h}(Q)\times n-\log_2\left(\frac{1}{2\varepsilon_{cor}\beta^2}\right),
\end{split}
\end{equation}
where n is the block size used for post processing, Q is the QBER, h is the binary entropy function truncated to unity at input values over 0.5, $\beta$ is optimised at $\varepsilon_{QKD}$/4, f$_{IR}$ is the efficiency of information reconciliation and $\varepsilon_{cor}$ is the probability with which the key is incorrect. 
The total measurement time increases with channel attenuation, although only 600~s are required at 30~dB channel loss to collect the required number of counts.

\begin{figure}[h!]
	\centering
	\includegraphics[width=0.98\linewidth]{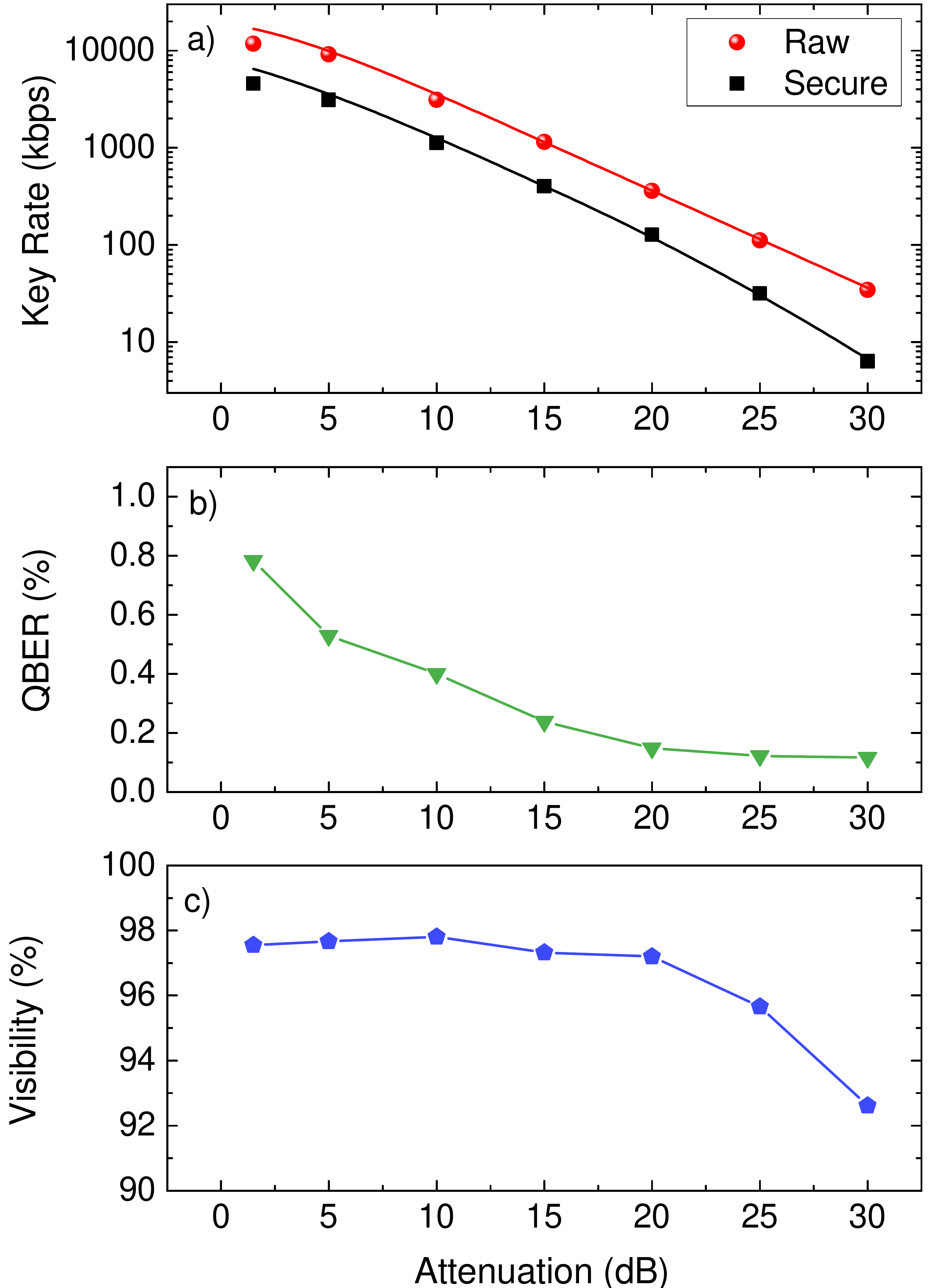}
	\caption{\textbf{Experimental (symbols) and simulated (lines) key rates and associated QBERs and visibilities.} COW protocol with a finite key-size analysis~\cite{Korzh_provably_2015}.}
	\label{fig:keyRateGraph}
\end{figure}

Fig.~\ref{fig:keyRateGraph}(a) shows the estimated secure key rate as a function of channel attenuation. 
This is the first time that megabit per second estimated key rates have been shown using the COW protocol.
These key rates are extracted in a finite key-size scenario, attaining 4.57~Mbit/s at 1.5~dB of attenuation. 
As the channel attenuation increases, the secure key rate decreases exponentially. 
At 20~dB of optical attenuation, equivalent to 100~km of ordinary optical fiber (0.2~dB/km loss), a secure key rate of 127.8~kbps is delivered. 
This rate is ten times higher than that measured by Korzh \textit{et al}~\cite{Korzh_provably_2015} using the COW protocol at similar attenuations. 
We attribute this enhanced performance to the lower QBER enabled by our source, alongside high efficiency detectors. 

We also plot the QBER and the interference visibility in Fig.~\ref{fig:keyRateGraph}(b). 
These parameters give a direct evaluation of the performance of our light source as a quantum transmitter. 
Because the single photon detectors have negligible DCRs, we do not expect a strong variation of QBER across the entire range of the channel attenuation. 
This is indeed the case for attenuations equal to and above 20~dB, where the QBER is measured at below 0.15~\%. 
This is low relative to other QKD protocols, which achieve QBERs of around 4~\% at similar distances~\cite{dixon_continuous_2010, Lucamarini_efficient_2013}. 
At the lowest channel attenuation of 1.5~dB, the QBER increases to 0.78~\%.  
We attribute this QBER increase to the deterioration of the timing jitter performance of the superconducting nanowire detectors at high count rates, where the jitter increases from 40~ps to 90~ps. 
We also increase the time-bin width on the digitizer at short distances to ensure all counts are measured. 
This deterioration causes an overlap between the detected time-bins, as shown in Fig.~\ref{fig:oscPatterns}(a), creating an ambiguity in the bit value of a photon. 

The interference visibility does not suffer from the time jitter deterioration because the count rate of SPD$_2$ is 30 times lower than SPD$_1$. 
As shown in Fig.~\ref{fig:oscPatterns}(b) and (c), the detection peaks are well separated from each other.  
We measure a visibility of 97.81~\% at 10~dB attenuation, illustrating high quality coherence transfer to the intensity-modulated pulses of the slave laser. 
This value is lower than the master laser visibility because the direct intensity modulation slightly weakens the indistinguishability among optical pulses due to the limited bandwidth of the slave laser. 
While our simulations show that improvement of the visibility would only entail a relatively small increase in the secure key rates, there is potential to reach far higher modulation rates using different slave laser diodes. 
Transmission at 10~Gbit/s has been shown in classical communications by using a gain-switched vertical-cavity surface-emitting laser with optical injection locking~\cite{Lin_beyond-bandwidth_2010}.

In summary, we have successfully demonstrated the suitability of a modulator-free QKD transmitter for the COW protocol. 
This system has produced estimated secure key rates between 4.57~Mbit/s to 6.38~kbit/s over equivalent distances of 7.5~km to 150~km. 
The lack of external modulators reduces both the system size and complexity. 
An exciting prospect opened up by this work is the potential for implementation in a multi-protocol network. 
Current work towards this has used bulky systems with a number of active components. 
The system presented in this work would enable a single transmitter to quickly switch between protocols depending on a client's requirement on bit-rate, distance or security. 
The wide range of functionalities offered by this transmitter, namely amplitude and phase modulation with on-demand phase randomization, mean that newly developed protocols could be easily adopted with firmware updates. 
An example of this would be an extension to the COW protocol that incorporates block-wise phase randomization to offer unconditional security, similar to work done for the DPS protocol~\cite{Tamaki_unconditional_2012}. 

\section*{Acknowledgment}
G. L. R. Acknowledges financial support via the EPSRC funded CDT in Integrated Photonic and Electronic Systems and Toshiba Research Europe Limited. 


\begin{thebibliography}{10}

\bibitem{gisin_quantum_2002}
N.~Gisin, G.~Ribordy, W.~Tittel, and H.~Zbinden, 
\newblock \textit{Rev. Mod. Phys.} \textbf{74} (1), 145--195 (2002).

\bibitem{Diamanti_practical_2016}
E.~Diamanti, H-K.~Lo, B.~Qi, and Z.~L.~Yuan, 
\newblock \textit{npj Quantum Information} \textbf{2}, 16025 (2016).

\bibitem{dixon_continuous_2010}
A.~R.~Dixon, Z.~L.~Yuan, J.~F.~Dynes, A.~W.~Sharpe, and A.~J.~Shields, 
\newblock \textit{Appl. Phys. Lett.} \textbf{96} (16), 161102 (2010).

\bibitem{Korzh_provably_2015}
B.~Korzh, C.C.W.~Lim, R.~Houlmann, N.~Gisin, M.J.~Li,
  D.~Nolan, B.~Sanguinetti, R.~Thew, and H.~Zbinden, 
\newblock \textit{Nat. Photonics} \textbf{9} (3), 163--168 (2015).

\bibitem{frohlich_long-distance_2017}
B.~Fr{ö}hlich, M.~Lucamarini, J.~F.~Dynes, L.~C.~Comandar, W.~W-S.~Tam, A.~Plews, A.~W.~Sharpe, Z.~L.~Yuan, and A.~J.~Shields, 
\newblock \textit{Optica} \textbf{4} (1), 163, (2017).

\bibitem{patel_quantum_2014}
K.~A.~Patel, J.~F.~Dynes, M.~Lucamarini, I.~Choi, A.~W.~Sharpe, Z.~L.~Yuan, R.~V.~Penty, and
  A.~J.~Shields, 
\newblock \textit{Appl. Phys. Lett.} \textbf{104} (5), 051123 (2014).

\bibitem{Qiu_quantum_2014}
J.~Qiu, 
\newblock \textit{Nature} \textbf{508}, 441, (2014).

\bibitem{Bennett_Quantum_1984}
C.~H.~ Bennett and G.~Brassard, 
\newblock in: International {Conference} on {Computer} {System} and
  {Signal} {Processing}, {IEEE}, 1984, pp. 175--179.

\bibitem{Scarani_security_2009}
V.~Scarani, H.~Bechmann-Pasquinucci, N.~J.~Cerf, M.~Du{š}ek,
  N.~Lütkenhaus, and M.~Peev, 
\newblock \textit{Rev. Mod. Phys.}  \textbf{81} (3), 1301 (2009).

\bibitem{sibson_integrated_2017}
P.~Sibson, M.~Godfrey, C.~Erven, S.~Miki, T.~Yamashita, M.~Fujiwara, M.~Sasaki, H.~Terai, M.~G.~Tanner, C.~M.~Natarajan, R.~H.~Hadfield, J.~O'Brien, and M.~G.~Thompson,
\newblock \textit{Nat. Comms.} \textbf{8}, 13984 (2017).

\bibitem{yuan_directly_2016}
Z.~L.~Yuan, B~Fr{ö}hlich, M~Lucamarini, G.~L.~Roberts, J.~F.~Dynes, and A.~J.~Shields, 
\newblock \textit{Phys. Rev. X} \textbf{6} (3), 031044 (2016).

\bibitem{Inoue_differential_2002}
K.~Inoue, E.~Waks, and Y.~Yamamoto, 
\newblock \textit{Phys. Rev. Lett.} \textbf{89} (3), 037902 (2002).

\bibitem{loudon_quantum_2000}
R.~Loudon, 
\newblock \textit{The quantum theory of light} (OUP Oxford, 2000).

\bibitem{liu_Modulator-free_2014}
Z.~Liu, J.~Kakande, B.~Kelly, J.~O'Carroll, R.~Phelan, D.~J.~Richardson, and R. Slavik,
\newblock \textit{Nat. Comms.} \textbf{5}, 5911 (2014).

\bibitem{He_amplification_2016}
J.~He, G.~Jin, B.~Liu and J.~Wang,
\newblock \textit{Opt. Lett.} \textbf{41}, 5724 (2016).

\bibitem{Comandar_quantum_2016}
L.~C.~Comandar, M.~Lucamarini, B.~Fr{ö}hlich, J.~F.~Dynes, A.~W.~Sharpe, S.W.-B.~Tam, Z.~L.~Yuan, R.~V.~Penty, and A.~J.~Shields,
\newblock \textit{Nat. Photonics.} \textbf{10}, 312 (2016).

\bibitem{Lucamarini_efficient_2013}
M.~Lucamarini, K.~A.~Patel, J.~F.~Dynes, B.~Fr{ö}hlich, A.~W.~Sharpe, A.~R.~Dixon, Z.~L.~Yuan, R.~V.~Penty, and A.~J.~Shields, 
\newblock \textit{Opt. Express} \textbf{21} (21), 24550 (2013).

\bibitem{Lin_beyond-bandwidth_2010}
C.~C.~Lin, Y.~C~Chi, H.~C.~Kuo, P.~C.~Peng, C.~J.~Chang-Hasnain, and G.~R.~Lin,
\newblock \textit{J. Lightw.Technol.} \textbf{29}, 830 (2011).

\bibitem{Tamaki_unconditional_2012}
K.~Tamaki, M.~Koashi, and G.~Kato, 
\newblock arXiv preprint arXiv:1208.1995 (2012).

\end{thebibliography}

\end{document}